# Scalable Group Management in Large-Scale Virtualized Clusters

Wei Zhou[1], Lei Wang[1], Dan Meng[1,2], Lin Yuan[1], Jianfeng Zhan[1]

[1] Institute of Computing Technology, Chinese Academy of Sciences
[2] Bureaus of High-Tech Research and Development, Chinese Academy of Sciences

*Abstract*—to save cost, recently more and more users choose to provision virtual machine resources in data centres. Maintaining a consistent member view is the foundation of reliable cluster managements, and it also raises several challenge issues for large-scale cluster systems deployed with virtual machines (which we call virtualized clusters). In this paper, we introduce our experiences in design and implementation of scalable member view management on large-scale virtualized clusters. Our research contributions are three-fold: 1) we propose a scalable and reliable management infrastructure that combines a peer-to-peer structure and a hierarchy structure to maintain a consistent member view in virtualized clusters; 2) we present a light-weight group membership algorithm that can reach the consistent member view within a single round of message exchange; and 3) we design and implement a scalable membership service that can provision virtual machines and maintain a consistent member view in virtualized clusters. Our work is verified on Dawning 5000A, which ranked No.10 of Top 500 super computers in November, 2008.

## 1. INTRODUCTION

To save cost, recently more and more users choose to provision resources at the granularity of *virtual machines* (*VMs*) in cluster systems, especially of data centres, and *virtualized clusters* emerges as promising platforms for both scientific computing and business services.

For a cluster system, *a member view* is the uniform list of members' real-time status (*running* or *crashed*) on cluster-wide. Maintaining *a consistent member view* is the foundation of reliable managements. For example, if a job scheduler, whose responsibility is assigning nodes for each job, gets an inconsistent member view, it will lead to incorrect resource assignments. For each cluster member, once a joining, leaving or crashing event happens, the member view of a cluster need to be updated with low overhead, which also raises a challenge issue for large-scale virtualized clusters. In our opinion, a qualified *member management system* (in short *management system*) has four requirements: first, the management system must maintain a consistent member view of the cluster. That is to say the management system *has and only has* a single uniform member view at any time; second, the management system is scalable without bottleneck when the cluster scale increases; third, the management system is reliable in that it can provide view managements with failure-resilience; fourth, the overhead of updating member view is low. That is to say once a view changing event (member's joining or crashing) occurs, the management system can be aware of the change and update the view immediately.

For virtualized clusters, providing effective and scalable member view management is more challenging than that of traditional cluster systems with the following reasons: first, members in virtualized clusters include both virtual machines (VM) and physical nodes, instead of only physical nodes in traditional cluster systems. In large-scale virtualized clusters, new virtual machines may be frequently created to join or destroyed to leave, and hence virtualized clusters are dynamic environments. While in traditional cluster systems, node's joining and leaving events are rare, and hence it is a relative static environment. Second, the scales of cluster systems in typical data centres (for example Google systems) have increased to more than ten thousands. Due to the fine-grained resource management, the number of VMs may be expanded to four or even ten times of that of physical nodes, and hence the scale of a virtualized cluster will increase to almost hundred thousand nodes. On this scale, failures are norms rather than exceptions.

In this paper, we focus on the scalable and effective group management (member view) for virtualized clusters. Previous work fails to resolve this issue in two ways: First, it lacks of a scalable management infrastructure, which has low overhead in updating members' view in large-scale virtualized clusters; second, it lacks of an inherent reliable mechanism in the management infrastructure.

In traditional cluster system, the group view management is an integrated part of system management or monitoring system [8] [9] [12] [13] [14]. Some work of [1] [2] [3] proposes a centralized management infrastructure that uses a single management server to monitor several management endpoints, which is not scalable in nature; some work [5] [6] presents a peer-to-peer management infrastructure of which several management endpoints constitute a peer-to-peer management structure, however complex group management protocols results in high overhead, which limits the scalability [4] [7].

In this paper, we propose a scalable group management service in virtualized clusters, which we call *SGMS*. SGMS has three major features: a) maintains a consistent member view of virtualized clusters continuously; b) provides the service of cluster member view; c) provisions virtual machines. The contributions of this paper can be concluded as follows:

1) We propose a scalable and reliable management infrastructure that merges a peer-to-peer structure and a hierarchy structure to maintain a consistent member view in virtualized clusters;

2) We present a light-weight group membership algorithm that can reach the consistent member view within a single round of message exchange;

3) We design and implement a scalable membership service that can provision virtual machines and maintain a consistent member view in large-scale virtualized clusters. Our work are verified on Dawning 5000A, which is ranked as top 10 of Top 500 super computers in November, 2008.

The paper is organized as follows: Section 2 presents the related work. Section 3 presents the scalable structure and Section 4 describes the consistent views of cluster member. Section 5 introduces the group membership algorithm. Section 6 is implementation. Section 7 is the extensive performance evaluation and Section 8 concludes the paper.

## 2. RELATED WORK

Most work designs view management system as an integrated part of system management or monitor management.

The work of [8] [9] adopts a hierarchy structure. For example, Ganglia [8], which is a scalable distributed monitoring system based on a hierarchy structure for clusters and Grids, adopts a tree structure among the head server and management servers. Ganglia [8] does not consider the reliability of the management servers.

The work of [4] [10] takes reliability and scalability into account within the hierarchy structure: Blue Eyes [4] adopts a multi-server scale-out structure to gain high scalability and reliability. It uses a binary tree to construct the relationships among head servers and management servers for scalability, and uses a k-level primary-backup logic ring structure for failure resilience. JOSHUA [10] implements symmetric active/active replications for head servers, and provides head servers with a virtually synchronous environment for continuous availability. It must add more spare services and need more nodes to deploy head servers. The work of [4] and [10] both needs redundancy components and requires more message exchanges, and they aim at decreasing clients' overheads for obtaining cluster member view while not considering the overhead of updating system view among head servers.

The work of [12] [13] [14] takes VM management into account within a hierarchy structure: EUCALYPTUS [12] [13] is an open-source implementation of Infrastructure as a Service system. Users can run and control entire VM instances deployed across a variety of physical resources, scaling from a single laptop to small Linux clusters. EUCALYPTUS uses a hierarchy structure. This work does not consider the member management system's performance and reliability in large-scale virtualized clusters.

Some previous efforts [5] [10] [15] adopt group membership idea for system management. HACMP [5] uses a peer-to-peer management structure based on a group membership algorithm to manage small-scale tight-coupled cluster system within the limit of 32 nodes; the work of [10] [15] uses a group communication service which can provide membership and reliable multicast services for group membership. JOSHUA [10] uses group communication to implement symmetric active/active replications for head servers. Galaxy management framework [15] focuses on servicing large-scale enterprise clusters. Galaxy can provide five different views of system, from weakly-consistent view of farm service membership to stronger-consistent view of cluster service membership. It completely adopts a peer-to-peer management structure, and is too complex to be suitable for view management in virtualized clusters.

## 3. SGMS STRUCTURE

SGMS is designed for virtualized cluster. Virtualized clusters are often comprised of large amount of physical nodes connected with commodity networks. Each physical node runs a virtual machine monitor and hosts multiple virtual machines which run several applications [16].

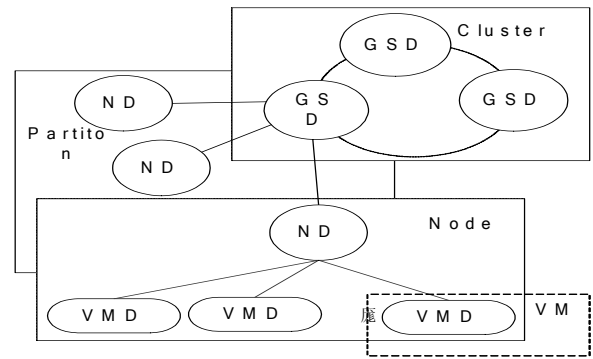

Fig.1. Architecture of SGMS

As shown in Fig.1, we divide a virtualized cluster into several partitions. Each partition has several nodes, and each node is deployed with several VMs. In practice, a virtualized cluster is divided into several partitions averagely. There are three components in SGMS: *group service daemon* (*GSD*), *node daemon* (ND) and *virtual machine daemon* (*VMD*). Each *GSD daemon*, *ND daemon* and *VMD daemon* is respectively deployed in each partition, each node and each virtual machine, responsible for managing a partition, a node and a virtual machine in the cluster. In our architecture, GSD is the management server, and ND is GSD's management endpoint. VMD is ND's management endpoint. In the rest of this section, we introduce the SGMS structure from two aspects: the structure in a partition and the structure among partitions, and then we conclude the characteristic of our structure.

### 3.1 The structure in a partition

As shown in Fig.2, We adopt a three-tier hierarchical structure in each partition: GSD-ND-VMD.

- *management structure*

We adopt *a three-tier hierarchy tree management structure* in a partition. GSD is the root node of the tree, and each ND in the partition is its leaf. Each VMD is the leaf of its parent ND. GSD is the single control point in each partition, which maintains the member view of the partition.

- *failure-resilience structure*

We adopt *a hierarchy failure-resilience structure* in each partition: each ND periodically sends heartbeats to its affiliated GSD in its managing partition, and GSD is responsible for detecting each ND and restarting the failed ND; ND detects each VMD on its hosting node and restarts the failed VMD.

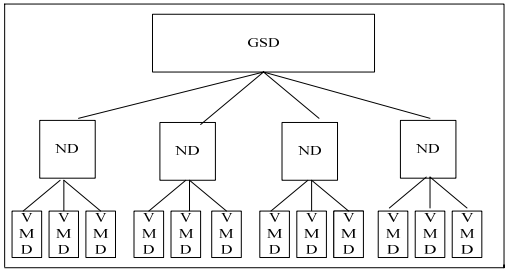

Fig.2. The structure in a partition.

### 3.2 structure among partitions

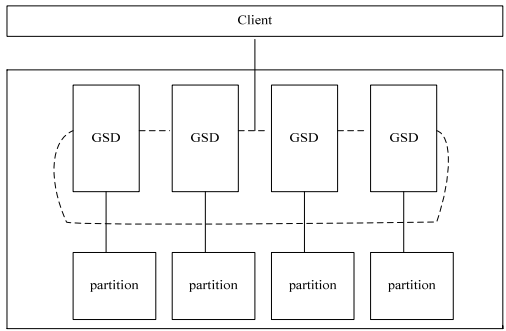

Fig.3. structure among partitions.

From Fig.3, we can see that all GSDs make up a group and maintain the membership of group. Each client can connect arbitrary GSD in the group to get the consistent group view in a cluster.

- *management structure*

Based on the group membership, the GSD group uses a *peer-to-peer* management structure. Each GSD manages its own partition. Once a client sends a command to a GSD, the connected GSD becomes the execution leader of this command, and will connect other GSDs in the current group membership list so as to make them execute the corresponding command, and finally the connected GSD returns the result to the client.

- *failure-resilience structure*

Based on the group membership, the GSD group uses a *peer-to-peer* failure-resilience structure. After one GSD leaves or crashes, other GSDs in the current group membership list will be aware of the change and then the GSD group performs failure-recovery operations.

### 3.3 The characteristic of the SGSM structure

In the SGMS structure, we can increase or decrease the number of partitions according to the scale of a cluster; the group membership algorithm of GSDs makes the SGSM structure work as a whole. On a basis of group membership, the *peer-to-peer* management structure of SGSM can maintain the consistent member view in a cluster and ensure load balance of each GSD in the group; the hierarchy failure-resilience structure in each partition and the *peer-to-peer* failure-resilience structure among partitions guarantee failure-resilience of the system.

In next section, we will introduce how to maintain the consistent member view in a cluster.

## 4. CONSISTENT MEMBER VIEWS IN CLUSTER

From the structure, we can see that a GSD maintains the member view in its own partition and the GSD group maintains the group membership among partitions.

### 4.1 Description of the View in SGMS

In order to maintain the member view in its own partition, a GSD must maintain *a VM view* in its own partition. The VM view is the real-time status of the VM list in the partition. The status includes: *running*, *crashed* and *suspended* (for virtual machine monitor providing the suspending function);

The status of a VM correlates to the status of its hosting physical node: If a physical node crashes, VMs residing in it will crash too; if the physical node leaves, VMs residing in it will be removed from the VM view too. GSD also maintains *the node view* in its own partition. The node view is the real-time status of the node list in the partition. The node status includes: *running* and *crashed*.

For the convenient expression, we formally describe the view and structure.

$GMS = SetOf(gsd)$, which represents the set of GSDs.
$GID = SetOf(gid)$, $gid$ is a positive integer and represents the ID of *gsd*.
$NDS = SetOf(node)$, which represents the set of physical nodes.
$NID = SetOf(nid)$, $nid$ is a positive integer and represents the ID of *node*.
$VMS = SeOf(vm)$, which representing the set of VMs.
$VID = SeOf(vid)$, $vid$ is a positive integer and represents the ID of a VM.
$ViewN = SeOf(NMember)$; $NMember = nid \times state$;
$nid \in NID$ and $state = Running \cup Crashed$.

*ViewV = SeOf (VMember); VMember=vid×state;*
*vid*∈*VID* and *state= Running*∪*Crashed*∪*Suspended*

*4.2 Implementing consistent member view*

- *Implementing consistent VM view*

Based on the three-tier hierarchy tree management structure, the GSD maintains the *ViewV* of its own partition. To achieve consistent view of *ViewV*, the GSD generates the *ViewV* of its own partition: the GSD controls NDs in its own partition, and each ND is responsible for the lifecycle management of VMs at its residing node. GSD will change the view as the VM's status changes; A ND monitors VMDs through heartbeats, if ND confirms the crash of a VM, it will inform the GSD to change the VM view.

- *Implementing consistent node view*

Based on the *hierarchy management structure*, a GSD maintains the *ViewN* of its own partition. To achieve consistent view of *ViewN*, the GSD monitors NDs through heartbeats. If the GSD finds a node is crashed, it will change the node view by setting the status of the node as *"crashed"*.

- *Implementation of the consistent view in cluster*

Based on the peer-to-peer management structure among partitions, SGMS provides an interface of accessing the group view for clients. The procedure for a client to get the VM view is as follows:

1) A client connects an arbitrary GSD and sends a request of getting the consistent VM view in the cluster.

2) The GSD connects other GSDs in the current group membership list and gets the VM views in other partitions.

3) The GSD sends the VM view in cluster to client.

## 5. GROUP MEMBERSHIP ALGORITHM

*5.1 Group Membership Problem*

The goal of group membership (GM) is to maintain a dynamic group of members and inform members about changes in the group [17] [18]. Changes in the group include: member's joining, leaving and crashing.

The goal of GM is maintaining the consistent list of the GSD group's current membership. We define the GSD group's current membership as *ViewG*.

**Define 5.1** *ViewG* : *ViewG = ViewId ×SetOf(GMember)*; *ViewId* is a natural number, represents the ID of *ViewG*; *GMember=gid×IP*, *gid*∈*GID* and *IP* is the IP address of GSD's reside node.

*5.2 Precondition of GM algorithm*

We list the preconditions of our GM algorithm, which include:

1) The communication between any two members is reliable and assures the FIFO order. The ordinary TCP protocol can assure this precondition.

2) The probability of simultaneous failures in a short time interval is almost zero.

3) No network partition occurs. In our algorithm, we assure that the cluster system has multi-redundant networks.

*5.3 Group specification and organization*

- *Group specification*

The membership satisfies three safety properties: *self inclusion, local monotonicity, primary component membership*; and one liveness property: *termination*.

*Self inclusion*, *local monotonicity* and *termination* property are the basic specifications for any group membership[18].

*Primary component membership* property[18] can ensure that the view of group is same for each member in the group at any time, implementing GSD group's *peer-to-peer* management and failure-resilience structure need this property.

**Property 5.1** *Self-Inclusion*

If *a gsd installs ViewG*, then *gid* ∈*ViewG*.

The self inclusion property requires the membership to maintain the unified view within the members in *ViewG*.

**Property 5.2** *Local Monotonicity*

If $ViewG_j$ is installed after $ViewG_i$ at $gsd_x$, then $ViewG_i$->$ViewId$ is strictly smaller than $ViewG_j$->$ViewId$.

The local monotonicity property requires the sequence of *ViewG* at each GSD to have monotonically increasing view identifiers.

**Property 5.3** *Primary Component Membership*

If $ViewG_j$ is immediately after $ViewG_i$, then $\exists gsd_x, gid_x \in ViewG_i \cup gid_x \in ViewG_j$

The Primary Component requires *ViewG* at each GSD is unified and total order.

**Property 5.4** *Termination*

If $gsd_x$ joins the group, then either $\exists ViewG_j, gid_x \in ViewG_j$ or gsdx eventually crashes;

if $\exists ViewG_j, gid_x \in ViewG_j$, and gsdx leaves or crashes, then $\exists ViewG_k, gid_x \notin ViewG_k$

The termination property requires the membership algorithm can achieve the convergence.

- *Group organization*

Group is organized as a unidirectional ring structure. We define the frontal and succeeding in the ring.

$gid_j = Front(gsd_i) -> (gsd_j = frontal(gsd_i)) \cap (gsd_i = succeeding(gsd_j))$

$$Front(gsd_i) = \begin{cases} \min(gid_j); gid_i < gid_j <= \max(gid) \\ 1; gid_i = \max(gid) \end{cases}$$

We assign three roles to the group: *Leader* which is the only one in the group and it is the member with *gid=1* at start-up; *Prince* which is the preceding one of *Leader* in the ring structure; *Member* which is the other in the group.

*5.4 Group bootstrap*

*Leader* is the coordinator in the group bootstrap. At start-up each member gets the static information and respective rank id from a system database. *Leader* initiates group creation through a two-phase commitment protocol. When a group is created, each member in the group has the same of *ViewG*.

### 5.5 Failure Detectors

We use heartbeat and timeout as the mechanism of failure detection in the group. Every member in a group will send heartbeats to its preceding one in the ring via all of its network interfaces periodically. If a member does not receive any heartbeats after timeout, it will doubt that its succeeding has been failed and send "Succeeding_failure" event.

### 5.6 GM algorithm

Types:
   *GMember=gid×IP; gid= Integer, IP=String;*
   *Set= SetOf(GMember); ViewId= Integer;*
   *ViewG  = ViewId × Set*

State:
| Type | Variable | Initial Value |
|---|---|---|
| *ViewG* | *ViewG* | *{1, Set }* |
| *Bool* | *send_ack* | *false* |
| *Bool* | *send_version* | *false* |
| *Bool* | *send_newViewG* | *false* |
| *Bool* | *send_crashreport* | *false* |
| *Bool* | *Judge_Result* | *false* |
| *Bool* | *recv_ack[Q], Q ∈ Set* | *false* |
| *Integer* | *rank* | *Leader∪Prince∪Member* |

Internal

*Chk_version*(sender_vid), *ViewId* sender_vid
eff: if (sender_VID< *ViewG* ->ViewId) then
    send_version<-true

*Judge_Failure***(**member**),** *GMember* member
eff: if (a member failures) then Judge_Result<- true

*Compensate* ()
eff:
  Forall *Q* in *ViewG -> Set*
   if (recv_ack[Q] = false) then
     *ViewG* ->ViewId++;
     (*ViewG* ->Set)<- (*ViewG* ->Set-Q) ;
     send_newViewG <-true;
   endif

*Wait_ack*()
eff:
  Forall *Q* in *ViewG -> Set*; recv_ack[Q] = false
  Forall *Q* in *ViewG -> Set*; Recv_ACK**(**sender**)**
endif

Input *Succeeding_failure* ()
eff:
 switch (*rank*)
 case *Leader*:
   *ViewG* ->ViewId++;
   (*ViewG -> Set*)<- (*ViewG -> Set-* Succeeding);
   send_newViewG <-true;
 case *Prince*:
   *ViewG* ->ViewId++;
   (*ViewG -> Set*)<- (*ViewG -> Set -* Succeeding)
   rank<-Leader; send_newViewG <-true
 case Member:
    send_crashreport<-true

Input *Recv_ACK*(sender), *GMember* sender
eff:
   recv_ack[sender] <-true
Input *Recv_CrashReport*(id,Sender,Crasher**),** *ViewId id, GMember* Sender, *GMember* Crasher
eff:
  if (Chk_version(id)=false) ∩(rank=*Leader*)
    ∩(Judge_Failure(Crasher)=true) then
    *ViewG* ->ViewId++;
    *ViewG* -> Set<- (*ViewG* -> Set -crashed);
    send_newViewG <-true;
  endif
Input *Recv_NewViewG***(***newV***),***ViewG newV*
eff:
  if (*newV* -> ViewId >*ViewG* ->ViewId) then
   *ViewG*<- *newV*
   if (my Front is Leader) then rank<-*Prince* endif
   send_ack<-true;
  endif
Input *Recv_Rejoining*(myinfo), *GMember* myinfo
eff:
 if ((myinfo.gid ∈ (*ViewG*->Set.gid)=false) ∩
   (rank=*Leader*)) then
  *ViewG* ->ViewId++;
  (*ViewG*->Set)<-(*ViewG*->Set+myinfo);
   send_newViewG<-true
 endif
Input *Recv_Joining*(myinfo), *IP* myinfo
eff:
  if (rank=*Leader*) then
   new_id<-max(gid:gid ∈ *ViewG*->Set.gid)+1
   newmember<-(*new_id×myinfo*)
   *ViewG* ->ViewId++;
   (*ViewG*->Set)<-(*ViewG*->Set + newmember)
   send_newViewG <-true
  endif
Input *Recv_ LeavingPropose*(*id,*Leaving), *ViewId id,* GMember Leaving
eff:
  if (Chk_version(id)=false) ∩(rank=*Leader*) then
   *ViewG* ->ViewId++;
   (*ViewG* ->Set)<-(*ViewG*->Set-Leaving)
   send_newViewG<-true;
  endif
Input *Recv_CurrenVersion*(id), *ViewId id*
eff: if (*ViewG* ->ViewId<id) then exit
Output Send_NewViewG
 pre: send_ newViewG =true
  eff:
   *Wait_ack*();
   *Compesate*()
Output *Send_CrashReport*(sender,crasher), *GMember* sender *, GMember* crasher
Output *Send_NewViewG (newV), ViewG* newV
Output *Send_LeavingPropose(sender),GMember sender*
Output *Send_Joining*(myinfo),IP myinfo
Output *Send_Rejoining(*myinfo), *GMember* myinfo
Output *Send_CurrenVersion*(id),*ViewId* id
Output *Send_ACK*(sender), *gid* sender

## 5.7 Algorithm Analysis

Our algorithm uses Leader dominate mechanism and Leader's Compensate mechanism (which correspond with the *Compensate()* function) to implement primary component membership with single round finished in partial synchrony environment. The time complexity of the algorithm is single round. The Leader dominate mechanism is that leader will deal with the event of joining, rejoining, leaving and succeeding failure, and take charge of the view updating. The *Compensate()* function is a recursion operation, it will not stop until the view of each member in the group is consistent.

From Fig.4 which *M* denotes *member*, *P* denotes *Prince* and *L* denotes *leader*, we can see that the view finishes its changes within a single round to without failures.

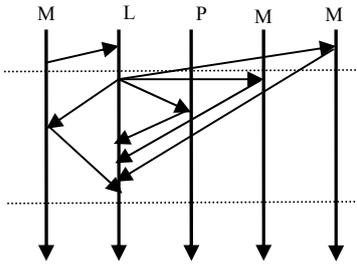

Fig.4. Single round of join operation

In the failure-occurred which is shown in Fig.5, we use leader's *Compensate()* function to implement single round finishing and primary component membership.

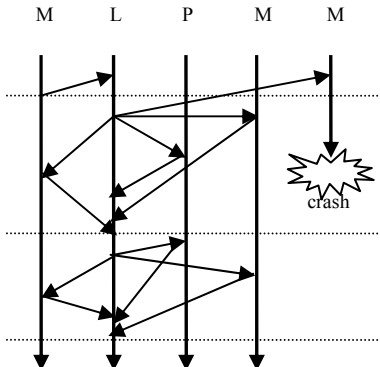

Fig.5. operation with other crashing

We don't deal with the condition that *Leader* and *Prince* fail simultaneously. This is because that the probability of simultaneous failure in short time interval is almost zero [19].

We implement primary component membership for GSD group's Peer-To-Peer management and failure-resilience structure.

## 6. IMPLEMENTATION

### 6.1 SGMS's components

There are three types of components in SGMS, which are GSD, ND and VMD.

For each partition, one GSD daemon is deployed. GSD is the manager of its resident partition, it manages VMs through ND and maintains the consistent view of VMs in its resident partition. It also maintains the membership of GSD group with others.

For each physical node, one ND daemon is deployed. ND is the basis management unit in each partition. It is responsible for VM lifecycle management at its resident node, notifying the health information of node to GSD by sending the heartbeat periodically.

For each VM, one VMD daemon is deployed. VMD is responsible for notifying the health information of virtual machine to ND by sending the heartbeat periodically.

### 6.2 SGMS's functions

SGMS can provide the consistent view of cluster members and dynamic provision of virtual machines for users. It also implements self failure-recovery.

#### 6.2.1 The functions for uses

SGMS's functions for uses include that: 1) providing the consistent view of cluster members for the user; 2) dynamic providing virtual machines for the user.

- *The interface*

The interface of providing the consistent view of cluster members is:

| Operations | Purpose |
|---|---|
| Get_VMs_state | Get a batch of VMs state. |
| Get_Cluster_state | Get VMs state in the cluster. |
| Inform_state | Call back the user that the VM states change. |

The interface of dynamic providing virtual machines is

| Operations | Purpose |
|---|---|
| Create_VMs | Create a batch of new VMs |
| Destroy_VMs | Destroy a batch of new VMs |
| VMs_Management | Manage the lifecycle of a batch of VMs, including: *start*, *shutdown*, *reboot*, *resize*, *hung* and *resume*. |

- *The implementation*

The functions for uses are provided by the GSD in the SGMS. We have describes the implementation of the consistent view of cluster members in section 4.

We implement dynamic providing virtual machine through the hierarchy management structure of GSD-ND. For the unique property of virtual machine provision, we design a map algorithm that each *vid* can be mapped to its host physical node's *nid* and each *nid* can be mapped to its host partition's *gid*; for the reliable property of virtual machine provision, we design a reliable VM provision protocol which is based on the transaction mechanism and logging rollback.

- *Self failure-recovery*

From Section 3, we know that the hierarchy failure-resilience structure in partition and GSD group's Peer-To-Peer failure-resilience among partitions implements the SGMS's failure-resilience. Now we describe the implement in details.

ND detects the failed VMD on its resided node with the heartbeat, and restart the failed VMD;GSD detects the failed ND in its resided partition through heartbeat, and restart the failed ND; when GSD fails, it will be recovered by GSD failure-recovery Protocol.

> 1] If GSD failure, its Frontal will do the failure-recovery operation.
> 2] Frontal GSD diagnoses failure GSD's failure reason.
> 3] If the failure reason is resided node crashed, GSD selects new node and restart the GSD; if the reason is process failed, GSD restart the GSD on the same node.
> 4] If the restart failed, GSD takeovers its Front's work for partition management.
> 5] If the restart success, the restarted GSD will execute the rejoin operation.

*6.3 system Implementation*

We use XEN as VMM (virtual machine monitor) to implement virtualized cluster, and we implement SGMS on the basis of ACE (http://www.cs.wustl.edu/~schmidt /ACE.html). ACE is a free, open-source object-oriented (OO) framework that implements many core patterns for concurrent communication softwares.

SGMS has the function of VM management, so we choose XEN-API (http://wiki.xensource.com/xenwiki/XenApi) which can provide operation APIs of Virtual Machine as the basis of configuring and controlling VM; we choose NFS as the global file system, SGMS can access and get the VM image; Mysql-4.0.13 is used as the database for the system to store information, and in order to guarantee the high availability of database, we choose Duplex machines HA solution (http://linux-ha.org/).

## 7. EVALUATION

In this section, we evaluate the performance of view management, the performance of GM algorithm, the performance of VM management, the performance of SGMS' Failure-recovery and the overload of SGMS.

The testbed is configured as follows: 136 X86-64 nodes(4 number of Quad-Core AMD Opteron CPU and 64G Memory), the operating system on each node is CentOS4.4 + Xen3.0.4, the VM is configured with two CPU, 4G Memory and CentOS4.4 operating system.

*7.1 The performance of maintaining consistent view*

The view management's performance is evaluated by the time spending on view updating and getting. It includes the time of client updating view and the time of client getting view.

- *the time of client getting view*

From the GSD unify interface protocol, we can know that the time of client getting the member view in cluster is only related to the number of GSDs in cluster. We divide 136 nodes into different number of partitions averagely and deploy 4 VMs at each physical node, use a test program to connect SGMS and test the performance of getting view.

Table.1 the overhead of client's getting view

| Number of partitions | Getting time (ms) |
|---|---|
| 2 | 4 |
| 4 | 8 |
| 8 | 15 |
| 12 | 23 |

From table.1, we can see that the time spending on getting view is about ten *milliseconds*.

- *the time of client updating view*

From Section 6 we know that *Inform_state( )* is the call back function for a client, if a client registers the *Inform_state( ) function* to SGMS, once a view changing event occurs, GSD will send information to the client immediately. And the time of client updating view in cluster is only related with number of nodes and VMs in each partition.

We use a test program to connect SGMS and register *Inform_state( )* function, then test the time of client updating view with different view changing event. The view changing event includes: nodes view changing in partition and VMs view changing in partition,

- *Nodes view changing*

Nodes view changing event includes: node joining, leaving and crashed. We divide 136 nodes into 8 partitions and deploy 4 VMs at each physical node. The number of nodes in the first partition is changed from 16 to 128. Then we generate the nodes view changing event, other partition distributes the rest nodes averagely. For test node crashed, we shutdown the node manually to simulate the crash; for test leave or join, we execute the node leave or join manually.

Table.2 the overhead of client's updating the nodes view

| Number of nodes | Node Joining (ms) | Node Leaving (ms) | Node Crashing (s) |
|---|---|---|---|
| 16 | 1.178 | 1.076 | N+0.266 |
| 32 | 1.279 | 1.173 | N+0.289 |
| 64 | 1.477 | 1.372 | N+0.321 |
| 128 | 1.83 | 1.71 | N+0.403 |

From table.2, we can see that the time spent on updating node joining and leaving is about milliseconds; the time spent on updating node failure is less than *(N+0.5)* seconds, where N is the timeout setting of heartbeat which may be from 1 to N second.

- *VMs view changing*

In SGMS, ND manages the VMs resided at its node and GSD manages the VMs resided at its partition through ND. The flow of VMs view changing is GSD sends VM manage command to ND, ND executes the command and report the updating information, ND also monitors the VM's heartbeat for crashing. We evaluate the VMs view changing in partition from two aspects: one is the time spent on client updating VM state; the other is the time spending on VM management operation. For the management operation is varying and we only test the performance of client updating VM state, the performance of VM management will show in next paragraph. The VM state updating includes: state updating with operation and state updating with VM crashed.

We divide 136 nodes into 8 partitions; first partition's node number is 128, the rest of nodes is averagely divided into others partitions; we choose the number of VMs resided at each physical node ranging from 1 to 8. To test state updating for VM crashing, we shutdown the VM manually to simulate the crash; for test state updating with operation, we execute the VM operator through using *VMs_Management () function,* which is defined in Section 6.2.

Table.3 the overhead of client's updating VMs view

| Number of Nodes | Number of VMs | VM operation (milliseconds) | VM crashing (seconds) |
|---|---|---|---|
| 128 | 1 | 1.75 | N+0.136 |
|  | 2 | 1.61 | N+0.141 |
|  | 4 | 1.67 | N+0.156 |
|  | 8 | 1.92 | N+0.166 |

The result of Table 3 shows that for normal operations the time spent in updating VM state is about milliseconds, while for VM crashing, the time spent in updating VM state is less than *(N+0.2)* seconds, where *N* is the timeout setting of heartbeat which may be from 1 to *n* second.

- *analysis*

From the implement mechanism of SGMS, we know that the system managing scale of SGMS is decided by the numbers of partition, the node numbers in one partition and the VM numbers in one node. From the above experiments, we infer that the node scale which SGMS can support is 128*12=1536 and the VM scale is 128*12*8=12288.

From the above experiments, we infer that the time of client getting view is less than 1s, the time spent on nodes or VM view changing is less than 3.5s (when the timeout setting of node heartbeat is 3s) under the condition of 1536 number of nodes and 12288 number of VMs.

### 7.2 The performance of GM algorithm

We evaluate the performance of GM algorithm by the time spent on updating membership list for join, rejoin, leave and crash. For test join, we divide 136 nodes into N+1 partitions averagely and construct the GSD group with first N partitions and then let the (N+1) partition join the group; For test leave, we divide 136 nodes into N partitions averagely and let Nth partition leave the group; For test crash, we divide 136 nodes into N partitions averagely and kill one GSD to simulate its crash; the rejoin will automate execute after GSD crash for the GSD failure-recovery we have implemented.

Table.4 the overhead of GSD updating membership list

| Number of partition | Updating list for join (ms) | Updating list for leave (ms) | Updating list for rejoin (ms) | Updating list for crash (ms) |
|---|---|---|---|---|
| 2 | 2.46 | 1.53 | 2.46 | 1.53 |
| 4 | 2.75 | 1.86 | 2.75 | 1.86 |
| 8 | 4.67 | 3.49 | 4.67 | 3.49 |
| 12 | 4.98 | 4.59 | 4.98 | 4.59 |

From table.4, we can see that the time spent on membership list updating is on the level of millisecond.

### 7.3 The performance of VM management

To evaluate the performance of VM management operation, we divide 136 nodes into 8 partitions averagely, and deploy 4 VMs at each physical node, and the total number of VMs is 544. We choose six VM operators to evaluate the performance of VM management.

Table.5 the performance of VM management

| Operation | The state after operation | time (seconds) |
|---|---|---|
| create+start | running | 3.765 |
| suspend | suspended | 6.992 |
| resume | running | 5.813 |
| shutdown | halted | 11.732 |
| resize | running | 0.391 |
| destroy | / | 0.1 |

From table.5, we can see that the average time of creating and starting a VM is 3.765 seconds, the average time of destroying a VM is only 0.1 seconds.

### 7.4 The performance of self failure-recovery

We evaluate the performance of self failure-recovery by the time spent on VMD, ND and GSD recover from failure. We divide 136 nodes into 8 partitions averagely and deploy 4 VMs at each physical node. Our Failure-recovery is based on heartbeat and timeout. We set heartbeat interval as 1s and timeout as 5s or 10s to simulate process failure by kill the process manually and node failure by shutdown the node manually. ND and VMD can only be deployed at its resident node and VM, when node failure occurs, SGMS will not recover them.

Table.6 the overhead of self failure-recovery

|  | Timeout Setting (seconds) | Failure reason | recovery time (seconds) |
|---|---|---|---|
| VMD | 10 | process failure | 12.443 |
|  | 5 | process failure | 7.543 |
| ND | 10 | process failure | 11.616 |
|  | 5 | process failure | 6.146 |
| GSD | 10 | process failure | 12.11 |
|  |  | node failure | 12.33 |
|  | 5 | process failure | 7.33 |
|  |  | node failure | 7.55 |

From table.6, we can see that the VMD, ND and GSD's failure can be detected and recover within (N+3) S, N is the timeout setting (the measurement unit is second).

### 7.5 Overload of SGMS

We evaluate the overload of service by testing SGMS components' CPU utilization percent and memory used in a day. From table.7 we can see that the GSD, ND and VMD use less of system resource.

Table.7 the overhead of SGMS

|  | CPU used (%) | MEM used (MB) |
|---|---|---|
| GSD | 0.0033 | 40 |
| ND | 0.0023 | 22 |
| VMD | 0.0023 | 2 |

### 8. CONCLUSION

In this paper, we have introduced the experience in designing and implementing a scalable view management service which calls SGMS to manage virtualized cluster system for Dawning 5000A, which ranked on the top 10 list of top 500 of super computer.

SGMS can maintain consistent member view continuously and provide the service of view and VM provision for other cluster management software. In SGMS we add group membership mechanism into hierarchy structure, constructing a reliable and scalable structure for large-scale virtualized cluster system. We design a group membership algorithm for the membership of group. The group membership algorithm designed for the partial synchrony environment is a primary component membership and can be finished with a single round.

From the performance evaluation, we can see that SGMS has reliable and scalable structure with low overhead of view updating in large-scale virtualized cluster environment.

### 9. ACKNOWLEDGEMENTS

This paper is supported by the NSFC programs (Grant No. 60703020 and Grant No. 60933003).